\documentclass{article}

\usepackage{microtype}
\usepackage{graphicx}
\usepackage{subcaption}
\usepackage{booktabs} % for professional tables
\usepackage{graphicx}
\usepackage{pifont}
\usepackage{multirow}
\usepackage{url} 
\usepackage[framemethod=TikZ]{mdframed}
\usepackage{listings}
\usepackage{xurl}
\usepackage{hyperref}
\usepackage[most]{tcolorbox}
\usetikzlibrary{shadows}

\usepackage{xcolor}
\usepackage[most]{tcolorbox}
\usepackage{xcolor}
\usepackage{listings}

\newtcolorbox{icmlbox}{
    enhanced,
    breakable,
    colback=gray!3,
    colframe=black!80,
    boxrule=1.2pt,
    arc=1.5mm,
    left=1.5mm,
    right=1.5mm,
    top=1mm,
    bottom=1mm,
    before skip=0.8em,
    after skip=0.8em,
    drop shadow={black!25},

    before upper={\ttfamily\small}
}

\newmdenv[
    backgroundcolor=gray!3,
    linecolor=black!80,
    linewidth=1.2pt,
    roundcorner=1.5mm,
    innerleftmargin=1.5mm,
    innerrightmargin=1.5mm,
    innertopmargin=1mm,
    innerbottommargin=1mm,
    skipabove=0.8em,
    skipbelow=0.8em,
    shadow=true,
    shadowsize=2pt,
    shadowcolor=black!25,
    frametitlefont=\ttfamily\small,
    settings={\ttfamily\small}
]{appendixpromptbox}

\usepackage[accepted]{icml2026}
\usepackage{amsmath}
\usepackage{amssymb}
\usepackage{mathtools}
\usepackage{amsthm}

\usepackage[capitalize,noabbrev]{cleveref}

\theoremstyle{plain}

\theoremstyle{definition}

\theoremstyle{remark}

\newcommand{\cmark}{\ding{51}}  % ✓
\newcommand{\xmark}{\ding{55}}  % ✗
\usepackage[textsize=tiny]{todonotes}

\icmltitlerunning{Inducing Overthink: Hierarchical Genetic Algorithm-based DoS Attack on Black-Box Large Language Reasoning Models}

\begin{document}

\twocolumn[
  \icmltitle{Inducing Overthink: Hierarchical Genetic Algorithm-based DoS Attack \\
  on Black-Box Large Language Reasoning Models}



  \icmlsetsymbol{equal}{*}
  \begin{icmlauthorlist}
    % \icmlauthor{Shuqiang Wang}{zju,equal}
    % \icmlauthor{Wei Cao}{zju,equal}
    \icmlauthor{Shuqiang Wang\textsuperscript{*}}{zju}
    \icmlauthor{Wei Cao\textsuperscript{*}}{zju}
    \icmlauthor{Jiaqi Weng}{comp}
    \icmlauthor{Jialing Tao}{comp}
    \icmlauthor{Licheng Pan}{zju}
    \icmlauthor{Hui Xue}{comp}
    \icmlauthor{Zhixuan Chu\textsuperscript{\textdagger}}{zju}
  \end{icmlauthorlist}
  \icmlaffiliation{zju}{The State Key Laboratory of Blockchain and Data Security, Zhejiang University}
  \icmlaffiliation{comp}{Alibaba Group}
  \icmlcorrespondingauthor{Zhixuan Chu}{zhixuanchu@zju.edu.cn}
  
  \icmlkeywords{Machine Learning, ICML}

  \vskip 0.3in
]
\printAffiliationsAndNotice{\icmlEqualContribution}



% no special notice (required even if empty)

\begin{abstract}
Large Reasoning Models (LRMs) are increasingly integrated into systems requiring reliable multi-step inference, yet this growing dependence exposes new vulnerabilities related to computational availability. In particular, LRMs exhibit a tendency to “overthink"—producing excessively long and redundant reasoning traces—when confronted with incomplete or logically inconsistent inputs. This behavior significantly increases inference latency and energy consumption, forming a potential vector for denial-of-service (DoS) style resource exhaustion.
In this work, we investigate this attack surface and propose an automated black-box framework that induces overthinking in LRMs by systematically perturbing the logical structure of input problems. Our method employs a hierarchical genetic algorithm (HGA) operating on structured problem decompositions, and optimizes a composite fitness function designed to maximize both response length and reflective overthinking markers.
Across four state-of-the-art reasoning models, the proposed method substantially amplifies output length, achieving up to a 26.1$\times$ increase on the MATH benchmark and consistently outperforming benign and manually crafted missing-premise baselines. We further demonstrate strong transferability, showing that adversarial inputs evolved using a small proxy model retain high effectiveness against large commercial LRMs. These findings highlight overthinking as a shared and exploitable vulnerability in modern reasoning systems, underscoring the need for more robust defenses.
\end{abstract}

\section{Introduction}

Large Language Models (\citealp{LLaMA}; \citealp{Training_language}; \citealp{Extracting}) have emerged as remarkably powerful AI tools, showing exceptional performance across a variety of real-world applications. Recently, Language Reasoning Models (LRMs), such as GPT-o3\yrcite{OpenO3}, DeepSeek-R1\yrcite{DeepSeek-AI} and Qwen3-Thinking \yrcite{Qwen3}, have further boosted the performance of LLMs in the domains of reasoning and programming, which are increasingly being embedded into systems requiring advanced cognitive capabilities. Professionals increasingly deploy these models to automate complex, multi-step tasks, thereby enjoying enhanced analytical power. As a result, critical systems, including enterprise software, scientific research platforms, and autonomous agents, increasingly integrate LRM capabilities to support sophisticated problem solving. In point of fact, 78\% of organizations reported using AI in 2024, a significant increase from 55\% the previous year \cite{stanford2025}. The global market for large language models was valued at over 5.6 billion US dollars in 2024 and is projected to grow to over 35.4 billion US dollars by 2030, demonstrating the rapid pace of adoption \cite{grandview2025}.\par
 
However, this growing dependence also introduces new attack surfaces and security risks to the system. LRMs take natural language prompts as input to perform complex computational and generative tasks. An attacker can exploit this interactive interface to inject specially crafted malicious prompts, which can induce the model to generate massive and redundant output. In large language models, output length is a primary driver of both computational energy consumption and inference latency\cite{inducing}. Therefore, this type of attack leads to severe service performance issues. Not only does this drastically drive up the economic costs for service operators, but it can also degrade service quality, effectively creating a Denial-of-Service (DoS) attack against other legitimate users.\par

Adversarial text attack techniques have made it possible to secretly inject malicious commands into Large Language Models(\citealp{autodan}; \citealp{zhang2025crabs}; \citealp{dong2025engorgio}). A significant limitation within the current landscape of computational resource attacks is the predominant focus on open-source, white-box models (\citealp{guo2021gradient};  \citealp{inducing}). These attack methodologies fundamentally rely on complete open access to the model itself, presupposing intimate knowledge of its internal architecture, parameters, and the ability to compute gradients directly. This prerequisite, however, renders such techniques largely inapplicable and impractical for contemporary commercial systems. \par

Moreover, existing research on DoS and energy-latency attacks is narrowly focused on general-purpose LLMs. This scope largely overlooks how the intricate reasoning chains within specialized reasoning LLMs can be uniquely exploited for significant cost amplification.

In this paper, we propose a novel, low-cost attack on black-box LRMs using an adversarial hierarchical genetic algorithm. The core idea is to exploit the inherent “overthinking” limitation of these models by automatically breaking down a problem's complete chain of thought. Our approach is inspired by two observations: (i) to secure reinforcement learning rewards, LRMs exhibit a tendency to engage in multiple rounds of repetitive thinking for any given question\cite{DeepSeek-AI}; and (ii) when confronted with problems that cannot be answered, LRMs often display a highly abnormal “overthinking” phenomenon\cite{chen2025overthinking}. By leveraging these characteristics, we first prompt the model to reveal its chain of thought for a problem, then automatically split it, and subsequently use a genetic algorithm to fracture the original logical chain, thereby achieving a computational resource attack. \par

This method can generate a series of adversarial attack texts that drastically increase the model's response length, leading to high energy consumption and latency, effectively achieving a DoS attack. Compared to the baseline case, the model's output length after applying this method can reach up to \textbf{26} times the original length. \par

\textbf{Contributions}. The contributions are as follows.
\begin{itemize}
    \item Propose automated black-box attack framework that uses a \textbf{genetic algorithm} on structured problem representations to systematically fracture logical chains and induce computational overthinking.
    \item Introduce a \textbf{composite fitness function} that guides the evolution by targeting not only output length but also 'overthink' markers. We demonstrate this composite function is significantly more effective at inducing resource exhaustion than optimizing for length alone.
    \item Establish the practical viability of the attack by demonstrating \textbf{high transferability}, showing that inputs evolved on a small, open-source proxy model retain high efficacy against large-scale, closed-source commercial systems.
\end{itemize}
\paragraph{Conflict of Interest Disclosure.}
The authors Jiaqi Weng, Jialing Tao and Hui Xue are employed by Alibaba Group, which lead the development of Model Qwen, which was among the models evaluated in this paper.
\section{Related Work}
\subsection{Overthinking Phenomenon and Its Harms}
Recent studies have highlighted \emph{overthinking} as an emergent failure mode of reasoning-oriented large language models (LLMs). 
Chen et al.~\yrcite{chen2025overthinking} demonstrate that o1-like models generate disproportionately long chain-of-thoughts (CoTs).
Fan et al.~\yrcite{fan2025missingpremise} further identify that when critical premises are absent, reasoning LLMs tend to self-entangle in redundant loops rather than abstaining. 
Cuadron et al.~\yrcite{cuadron2025danger} extend this line of inquiry to \emph{agentic environments} such as software engineering tasks. They reveal that models prefer internal simulation over environmental interaction, leading to \emph{analysis paralysis}, \emph{rogue actions}, or \emph{premature disengagement}. 
Dang et al.~\yrcite{Ren2025universal} reveal for the first time that overthinking in reasoning models may stem from their internal bias towards input texts.
Collectively, these works establish that overthinking not only reduces efficiency but also degrades effectiveness across diverse scenarios.

\subsection{Denial-of-Service and Energy Attacks on LLMs}
Parallel to the study of overthinking, another research thread explores how adversaries can deliberately \emph{inflate inference costs} via denial-of-service (DoS) or energy-latency attacks. 
Shumailov et al.~\yrcite{shumailov2021sponge} introduce \emph{sponge examples}, which maximally stress NLP models’ computational dimensions, increasing energy usage and latency by up to $6000\times$ in real-world translation services. 
Geiping et al.~\yrcite{geiping2024coercing} systematize adversarial objectives for LLMs and present a \emph{DoS attack} that suppresses the EOS token. 
Gao et al.~\yrcite{inducing} identified a direct proportionality between the output length of Large Language Models and both energy consumption and latency. Consequently, they developed delayed EOS loss, uncertainty loss, and token diversity loss to explicitly incentivize the model to engage in over-generation. 
Dong et al.~\yrcite{dong2025engorgio} propose \emph{Engorgio prompts}, optimized in embedding space to suppress EOS and achieve outputs near the context length limit. 
Zhang et al.~\yrcite{zhang2025crabs} introduce \emph{Crabs (AutoDoS)}, that builds a \emph{DoS Attack Tree} and uses \emph{Length Trojan} strategies to expand prompts. 
Kumar et al.~\yrcite{kumar2025overthink} propose the \emph{OverThink attack}, which exploits reasoning models’ vulnerability to decoy tasks thereby inflating reasoning tokens making detection particularly difficult. 
Zhang et al.~\yrcite{zhang2025one} introduce \emph{Deadlock}, which drives models into near-infinite “keep thinking” loops by repeatedly triggering transitional tokens such as “Wait” or “But” at the end of reasoning steps, effectively inflating computational effort. 
Li et al.~\yrcite{li2025pot} propose \emph{POT}, which leverages an external LLM to iteratively optimize prompts, automatically generating semantically natural and stealthy guidance phrases that encourage excessively long reasoning. However, it lacks the ability to systematically explore reasoning-aware perturbations at the problem level. 
More recently, Li et al.~\yrcite{li2025thinktrap} develop \emph{ThinkTrap}, mapping discrete prompts into a continuous surrogate space and performing derivative-free optimization in a low-dimensional subspace; while effective in black-box DoS scenarios, the resulting optimized prompts are generally less interpretable and produce attack inputs with limited readability.
These works collectively show that LLMs’ resource consumption can be deliberately amplified, yet they mainly focus on \emph{generic LLMs or shallow prompt manipulations}, with limited attention to reasoning-oriented models’ unique vulnerabilities.
\begin{figure*}[t]
    \centering
    \includegraphics[width=0.95\textwidth,keepaspectratio]{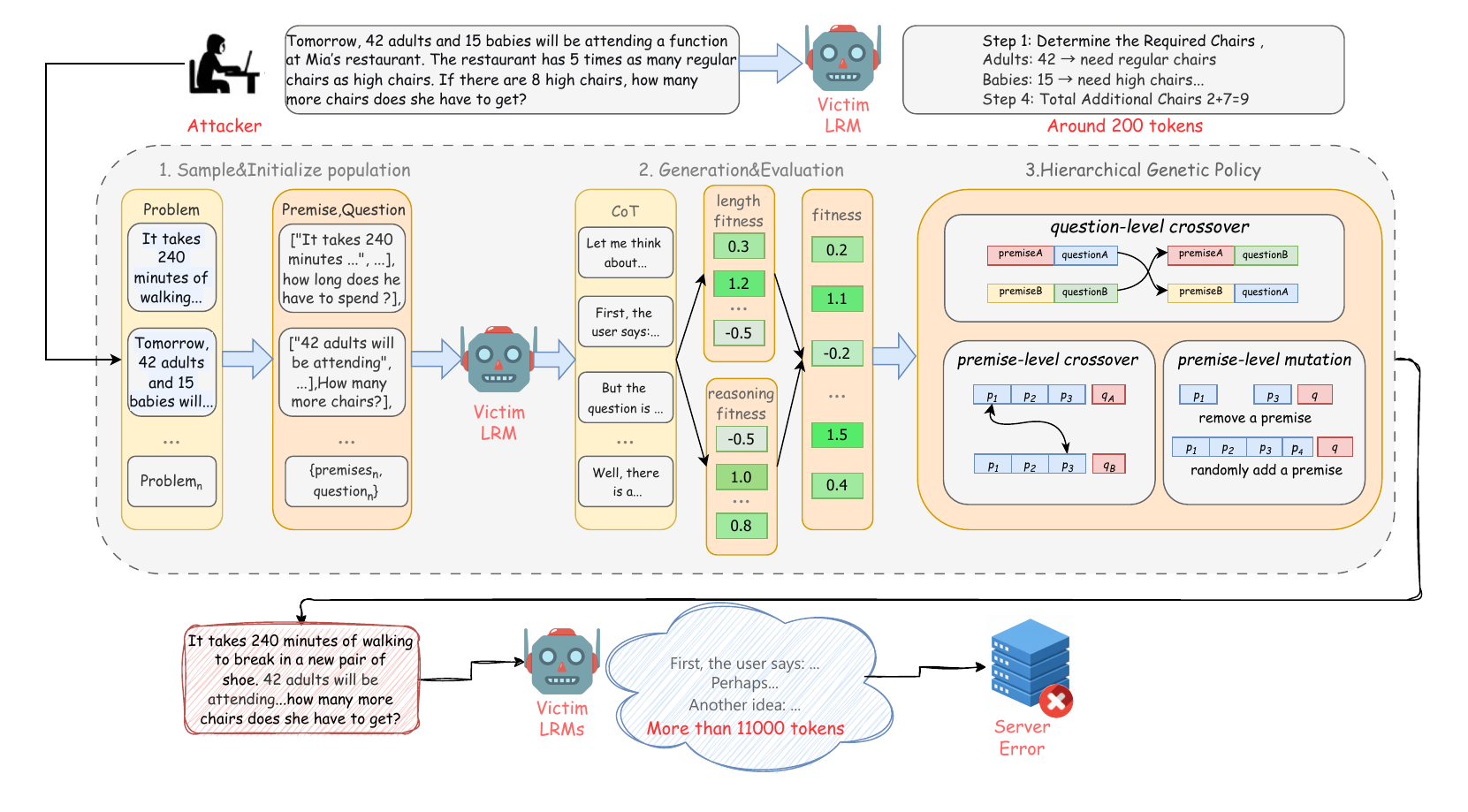}
    \caption{Overview of our proposed adversarial framework. 
The process starts with sampling from a dataset and initializing them to structured format, followed by fitness evaluation against the victim LRM. 
Through genetic policies including question-level and premise-level crossover and mutation, new inputs are generated to induce long and complex reasoning in large reasoning models.}
    \label{fig:overview}
\end{figure*}
\subsection{Motivation}
Building on these two threads, our work addresses critical gaps at their intersection: 
\textbf{Scope limitation}: Existing DoS and energy-latency attacks mostly target \emph{general-purpose LLMs}, with little exploration of how \emph{reasoning chains} in specialized reasoning LLMs can be exploited for cost amplification.  

\textbf{Attack quality}: Previous white-box methods such as gradient-based prompt optimization (GCG \cite{zou2023universal}; Engorgio \cite{dong2025engorgio}) often produce adversarial inputs that lack logical or semantic coherence, making them less effective at `\emph{extending reasoning chains} and more easily detectable.  

 \textbf{Automation gap}: Although Fan et al.~\yrcite{fan2025missingpremise} highlight that missing premises can exacerbate overthinking, their construction is manual and not adversarially optimized. In contrast, we propose a \emph{genetic algorithm--based automated pipeline} that transforms normal well-posed questions into adversarial ones, systematically inducing \emph{overthinking behaviors} in reasoning LLMs.

\begin{table}[t]
\centering
\setlength{\tabcolsep}{3pt}
\renewcommand{\arraystretch}{1.1}
\caption{Comparison between representative LLM-DoS or cost-amplification attacks and our method.}
\label{tab:comparison}
\begin{tabular}{@{}p{1.55cm}ccc p{1cm}@{}}
\toprule
\textbf{Method} & \textbf{Black-box} & \textbf{Reasoning-aware} & \textbf{Automated} \\
\midrule
GCG & \xmark & \xmark & \cmark  \\[2pt]
AutoDoS & \cmark & \xmark & \cmark  \\[2pt]
OverThink & \cmark & \cmark & \xmark  \\[2pt]
\textbf{Ours} & \textbf{\cmark} & \textbf{\cmark} & \textbf{\cmark}  \\
\bottomrule
\end{tabular}
\end{table}

\noindent
\textbf{Discussion.}
As summarized in Table~\ref{tab:comparison}, existing DoS-style attacks differ significantly in their assumptions and targets. 
GCG~\cite{zou2023universal} performs white-box gradient-based suffix optimization, but its generated adversarial inputs lack semantic structure and are not reasoning-aware.
Crabs (AutoDoS)~\cite{zhang2025crabs} achieves black-box scalability but primarily manipulates surface-level prompt length rather than reasoning processes.
OverThink~\cite{kumar2025overthink} focuses on reasoning token amplification via injected decoy tasks, yet requires manual prompt design and lacks automation.
In contrast, our proposed \textit{Overthinking Induction} attack simultaneously satisfies all three desiderata—\textbf{black-box applicability}, \textbf{reasoning awareness}, and \textbf{automated adversarial input generation}—by leveraging a genetic algorithm that evolves semantically valid reasoning problems to systematically induce overthinking in reasoning LLMs.HGA automates the discovery and evolution of the adversarial logical structures themselves.

\section{Methodology}

\label{sec:methodology}
In this work, we propose a hierarchical genetic algorithm (HGA)–based framework to generate adversarial problem formulations that maximize the cognitive load on large reasoning models (LRMs). The underlying hypothesis is that LRMs, when confronted with logically inconsistent or structurally perturbed problems, tend to exhibit overthinking behavior, resulting in longer responses. Our methodology is structured into four key stages: representation, fitness evaluation, selection, and genetic operators.

\subsection{Problem Representation}
In order to analyze and manipulate the reasoning behavior of large reasoning models (LRMs), it is necessary to represent each input problem in a form that explicitly exposes its internal logical structure. 
Rather than treating a problem as an opaque text sequence, we decompose it into its fundamental reasoning components, 
allowing fine-grained control over how individual elements contribute to the overall reasoning chain. 
This structured representation not only enables systematic perturbations and recombinations of problem elements 
but also provides a clear foundation for defining crossover and mutation operations within the genetic framework. \par

We treat each user input problem as a structured set composed of several premises and a final question. 
Formally, an individual is represented as:
\begin{equation}
    x = (\mathbf{P}, q),
\end{equation}
where
    $\mathbf{P} = [p_{1}, p_{2}, \dots, p_{n}]$ is a set of \emph{premises},
    $q$ is the \emph{final question}.
Thus, the space of all valid problem instances is defined as
\begin{equation}
    \mathcal{X}
    =
    \left\{
        (\mathbf{P}, q)
        \;\middle|\;
        \mathbf{P} \subseteq \mathcal{S},\; q \in \mathcal{S}
    \right\},
\end{equation}
where $\mathcal{S}$ denotes the set of all valid textual statements.\par

A population in generation $t$ consists of $N$ individuals:
\begin{equation}
    \mathcal{P}^{(t)} = \left\{ x^{(t)}_1, x^{(t)}_2, \dots, x^{(t)}_N \right\}.
\end{equation}
Each individual $x_i^{(t)}$ is initialized by decomposing a dataset problem into its premise list and question using the template in Appendix~\ref{appendix:decompose_template}.

\subsection{Fitness Function}
A key challenge in inducing overthinking behavior lies in defining an objective metric that can reliably capture the cognitive load and reflective tendencies of large reasoning models (LRMs). 
Beyond measuring the length of the chain of thought (CoT), overthinking may also manifest through linguistic cues such as hesitation, revision, or self-correction. 
Based on this observation, we design a composite fitness function that jointly evaluates both the depth and the reflectiveness of reasoning. 
Specifically, we quantify two complementary aspects of overthinking: 
(1) the verbosity of the reasoning chain, and (2) the occurrence of explicit overthinking markers in the model’s output. 
The combination of these two indicators enables the algorithm to distinguish between mere verbosity and genuine overthinking behavior, thereby aligning the optimization objective with the intended attack mechanism. \par
For each individual ${x}$, we define: \par
\textbf{Verbosity Score}.
Let $R(x)$ denote the chain-of-thought (CoT) response generated by the target LRM for input $x$.  
The verbosity score measures the overall length of the reasoning trace:
\begin{equation}
    score_{1}(x) = |R(x)|,
\end{equation}
where $|R(x)|$ denotes the total number of tokens in the CoT output.  
This term captures the degree to which the model produces extended or unnecessarily long reasoning chains.

\textbf{Reflective Marker Score}.
Overthinking often manifests through explicit hesitation or self-correction markers.  
Let $\mathcal{V}$ be a predefined vocabulary of overthinking markers.  
The reflective marker score is defined as
\begin{equation}
    score_{2}(x) 
    = 
    \sum_{w \in \mathcal{V}}
    \operatorname{Count}(w, R(x)),
\end{equation}
where $\operatorname{Count}(w, R(x))$ denotes the frequency of token $w$ appearing in the generated reasoning trace.  
This term quantifies explicit signs of hesitation or re-evaluation.\cite{InnateReasoning}

Both $score_1$ and $score_2$ are normalized within each generation using \textit{z-score} normalization. For a given score $score_i$ of individual $i$, the normalized value $\widehat{score}_i$ is defined as:
% \begin{equation}
%     \hat{s}_i
%     =
%     \frac{s_i - \mu}{\sigma},
% \end{equation}
\begin{equation}
\label{eq:z_score_norm}
\widehat{score}_{i}(x) = \frac{score_{i}(x) - \mathbb{E}_{x \in \mathcal{P}^{(t)}}[score_{i}(x)]}{\sqrt{\text{Var}_{x \in \mathcal{P}^{(t)}}[score_{i}(x)]}}
\end{equation}

The final fitness is then computed as a weighted combination of the two normalized scores:
\begin{equation}
    f(x) = \alpha \cdot \widehat{score}_{1}(x)
    + (1-\alpha) \cdot \widehat{score}_{2}(x)
\end{equation}
where $\alpha \in [0,1]$ balances verbosity against explicit overthinking behavior.

\subsection{Selection Strategy}
We employ a hybrid selection strategy that combines elitism with roulette-wheel selection: \par
\textbf{Elitism.} The top-performing individuals with the highest fitness scores are directly carried over to the next generation to ensure that the best solutions are not lost.\par
\textbf{Roulette-wheel selection.} The remaining parents are chosen probabilistically based on their relative fitness values. This balances exploitation of high-fitness individuals with the exploration of diverse candidates.

\subsection{Hierarchical Genetic Policies}
The genetic operators in our framework are designed to explore the space of logical perturbations in a structured yet stochastic manner. 
Unlike conventional genetic algorithms where crossover and mutation mainly focus on feature recombination or parameter search, 
our operators directly manipulate the logical structure of problems to induce reasoning disturbances in LRMs. 
These operations do not aim to preserve semantic or syntactic coherence; 
instead, they intentionally modify or distort the logical dependencies between premises and questions. 
Such manipulations create perturbed problem instances that remain interpretable to the model input pipeline 
but deviate from natural human reasoning patterns, thereby providing an effective mechanism 
to probe the model’s sensitivity to inconsistent or ill-structured logic. 
Through this process, the algorithm systematically searches for inputs that maximize the model’s reasoning load and overthinking tendency.\par
We define two operators, crossover and mutation, each acting at a different level of the problem structure.

    \subsubsection*{Crossover} With crossover probability \( p_{c} \), two parent individuals
    \[
    x_{A} = (\mathbf{P}_{A}, q_{A}), 
    \qquad
    x_{B} = (\mathbf{P}_{B}, q_{B})
    \]
    are combined to generate offspring.  
    We employ two levels of crossover:\par
        \textit{Question-level crossover:} This operator exchanges the final questions of the two parents:
\begin{equation}
\begin{aligned}
x_{C} &= (\mathbf{P}_{A},\; q_{B}), \\
x_{D} &= (\mathbf{P}_{B},\; q_{A}).
\end{aligned}
\end{equation}
This effectively mismatches a premise set with an unrelated question,  
creating logically fractured problem instances.\par
         \textit{Premise-level crossover:} Let \( k_A \sim \mathrm{Uniform}\{1,\dots,|\mathbf{P}_A|\} \)  
and \( k_B \sim \mathrm{Uniform}\{1,\dots,|\mathbf{P}_B|\} \)  be randomly sampled premise indices.  
The operator swaps the selected premises:
\begin{equation}
\begin{aligned}
x_{C} &= 
\big(
   \mathbf{P}_{A} \setminus \{p^{A}_{k_A}\}
   \; \cup \;
   \{p^{B}_{k_B}\},
   \;
   q_{A}
\big), \\[4pt]
x_{D} &= 
\big(
   \mathbf{P}_{B} \setminus \{p^{B}_{k_B}\}
   \; \cup \;
   \{p^{A}_{k_A}\},
   \;
   q_{B}
\big).
\end{aligned}
\end{equation}
In each crossover, the question-level crossover occurs with a probability of \( p_{qc} \), while the premise-level crossover occurs with a probability of \(1-p_{qc} \). This produces cross-context premise combinations that disrupt the original reasoning chain.

\subsubsection*{Mutation}
Mutation is applied to an individual \( x = (\mathbf{P}, q) \) with probability \( p_{m} \).
We consider two mutation modes, both operating at the premise level.

\paragraph{Premise Deletion}
A randomly sampled premise index  
\( k \sim \mathrm{Uniform}\{1,\dots,|\mathbf{P}|\} \)  
is removed:
\begin{equation}
\mathrm{Delete}(x)
=
(\mathbf{P} \setminus \{p_{k}\},\; q).
\end{equation}

\paragraph{Premise Addition}
Given another individual  
\( x' = (\mathbf{P}', q') \)  
sampled from the population,  
a random premise \( p'_{j} \in \mathbf{P}' \) is inserted:
\begin{equation}
\mathrm{Add}(x, x')
=
(\mathbf{P} \cup \{p'_{j}\},\; q).
\end{equation}
This operation increases logical inconsistency by incorporating unrelated external premises.

\subsection{Evolutionary Process}
The algorithm proceeds for a predefined number of generations. In each generation: Compute the fitness scores of all individuals.Select parents via elitism and roulette-wheel selection.Apply crossover and mutation to generate offspring.Form the next generation by combining elites and offspring.

When the termination criterion is met, the individual with the highest fitness score across all generations is identified as the optimal adversarial problem, i.e., the input that triggers the longest reasoning response.

\begin{algorithm}[ht]
\caption{Overthinking Induction}
\label{alg:GA}
\begin{algorithmic}[1]
\STATE Initialize population $P$ with structured problems from dataset
\FOR{generation $=1$ to $G$}
    \STATE \textbf{Fitness Evaluation:} For each individual $x \in P$, compute $f(x)$ as the fitness.
    \STATE \textbf{Selection:}
    Retain top-$N$ elites and
    Select remaining parents via roulette-wheel sampling
    \STATE \textbf{Crossover (with probability $p_c$):}\par
        \textit{Question-level crossover:} Exchange questions between two parents \par
        \textit{Premise-level crossover:} Randomly exchange one premise between two parents
    \STATE \textbf{Mutation (with probability $p_m$):}\par
        \textit{Deletion:} Randomly remove one premise from the offspring\par
        \textit{Addition:} Insert a random premise borrowed from another individual
    \STATE Form new population $P$ from elites and offspring
\ENDFOR
\STATE Return best individual with maximum fitness
\end{algorithmic}
\end{algorithm}

\section{Evaluation}
\label{sec:evaluation}
\subsection{Experimental Setup}
\textbf{Models.}
We evaluate our attack method on four representative large reasoning models:
DeepSeek-R1 (671B)\yrcite{DeepSeek-AI}, Qwen3-Thinking\yrcite{Qwen3}, GPT-o3\yrcite{OpenO3} and Gemini-2.5-Flash\yrcite{Gemini}.
All models are accessed through API interfaces under identical temperature and decoding parameters.
The attacks are conducted in a strict black-box setting without access to model gradients or internal weights.\par

\textbf{Baselines.}
We compare our adversarially generated inputs against two baseline datasets that represent distinct levels of logical integrity.  
(1) \textit{Original Dataset (Clean)} — the unmodified reasoning problems from three clean datasets. 
(2) \textit{Missing Premise Dataset}\cite{fan2025missingpremise} — A synthetic benchmark specifically designed for manually created overthinking caused by missing logical premises.
Each problem in this dataset omits one or more essential conditions, leading models to engage in prolonged and redundant reasoning loops. \par
These baselines allow us to isolate the effect of logical incompleteness on reasoning amplification.  

\textbf{Datasets.}
We construct our evaluation dataset from three reasoning-oriented benchmarks: GSM8K\cite{cobbe2021GSM8K}, SVAMP\cite{patel-etal-2021-nlp} and MATH \cite{hendrycksmath2021}(competition-level problems).
These datasets are particularly suitable as they naturally contain multi-step logical dependencies and clear ground-truth answers. \par
Given that the MIP dataset does not encompass all questions from the initial dataset, we filtered the base dataset to retain only those questions present in the MIP collection. and use LRM to decompose them into structured representations of premises and questions to form the initial population for the genetic algorithm. In this experiment, we used Qwen3-Thinking to segment the problem\par

\textbf{Attack Settings.}
Our attack follows the genetic algorithm framework described in Section 3, aiming to maximize the composite fitness function that reflects both response verbosity and reflective overthinking.
The population size is set to $N = 10$, and the algorithm evolves for $G = 5$ generations.
Crossover and mutation probabilities are $p\_c = 0.8$ and $p\_m = 0.2$, respectively.
Each fitness evaluation requires querying the victim model once and measuring its reasoning response.
For fairness, all models are evaluated under the same prompt budget and sampling temperature.
The total query budget per model is approximately 60 evaluations.
The attack strictly adheres to a black-box, query-based threat model, assuming the adversary can only submit textual inputs and observe textual outputs.\par

\begin{table}[t]
\centering
\caption{Baseline datasets and comparison with our method.}
\resizebox{\columnwidth}{!}{
\begin{tabular}{lcccc}
\toprule
Dataset / Method & Perturbation & Construction & Automated & Token Gain \\
\midrule
Original & None & Human-authored & \xmark & 1.0× \\
MIP & Missing Premise & Manual synthetic  & \xmark & 3.2× \\
\textbf{Ours} & Logical perturbation & Automated (GA) & \cmark & \textbf{26.1×} \\
\bottomrule
\end{tabular}}
\label{tab:baselines}
\end{table}
\par
\textbf{Metrics.} The evaluation framework is designed to quantify the computational cost inflicted on the target LRM. In modern transformer-based inference systems, key performance indicators such as response latency and energy consumption are directly proportional to the number of tokens generated by the model. Therefore, the output token length is selected as the principal proxy metric to measure the effectiveness of the overthinking-induction attack.To provide a comprehensive and robust assessment, two distinct measures are employed for each dataset: \textbf{Average Output Token Length (Avg-len):} To reduce biases that may arise from stochasticity in the inference process or other inherent model limitations, the average response length is computed across all test cases. This metric provides a balanced and representative measure of the attack framework's general efficacy. \textbf{Maximum Output Token Length (Max-len):} Considering the extreme requirements for a successful Denial-of-Service (DoS) attack, the adversary's goal is to trigger a worst-case scenario that exhausts system resources. The single longest response generated for each dataset is therefore also recorded. This metric serves as the standard for judging the final severity of the attack and its potential to saturate the model's maximum context length. Both the average and maximum output token lengths correspond to the results obtained during the final iteration of the HGA algorithm.\par
\subsection{Results}

\begin{table*}[t]
\centering
\caption{Combined results of the DoS Attack against modern thinking LLMs}
\resizebox{\textwidth}{!}{
\begin{tabular}{l l l l l l l l l l}
\toprule
Dataset & Model & \multicolumn{2}{c}{\textbf{DeepSeek-R1}} & \multicolumn{2}{c}{\textbf{Qwen3-Thinking}} & \multicolumn{2}{c}{\textbf{GPT-o3}} & \multicolumn{2}{c}{\textbf{Gemini-2.5-Flash}} \\
& & Max-len & Avg-len & Max-len & Avg-len & Max-len & Avg-len & Max-len & Avg-len \\
\midrule

% Use \multirow{3}{*}{...} to merge 3 rows in the first column
\multirow{3}{*}{SVAMP} & BASE & 360 & 202 & 1153 & 634 & 773 & 239 & 871 & 455 \\
& MIP & 4098 & 2319 & 3744 & 2231 & 906 & 620 & 683 & 380 \\
& Ours & 7814 & 3589 & 7906 & 5447 & \textbf{6562} & 3346 & 941 & 594 \\
\midrule
% --- GSM8K data ---
\multirow{3}{*}{GSM8K} & BASE & 500 & 343 & 1004 & 765 & 668 & 367 & 625 & 530 \\
& MIP & 6035 & 3093 & 3619 & 1521 & 2013 & 1038 & 636 & 409 \\
& Ours & \textbf{9068} & 4121 & 6344 & 4082 & 3089 & 1405 & 1209 & 987 \\
\midrule
% --- MATH500 data ---
\multirow{3}{*}{MATH} & BASE & 467 & 355 & 9016 & 3618 & 1039 & 416 & 6775 & 2889 \\
& MIP & 6769 & 4323 & 17064 & 7184 & 1502 & 1177 & 15523 & 8043 \\
& Ours & 12206 & 8817 & 22303 & \textbf{13007} & 2198 & 1618 & \textbf{18011} & 12147\\
\bottomrule
\end{tabular}}
\label{tab:combined_baselines}
\end{table*}
\begin{figure}[t]
    \centering
    \includegraphics[width=0.95\columnwidth,keepaspectratio]{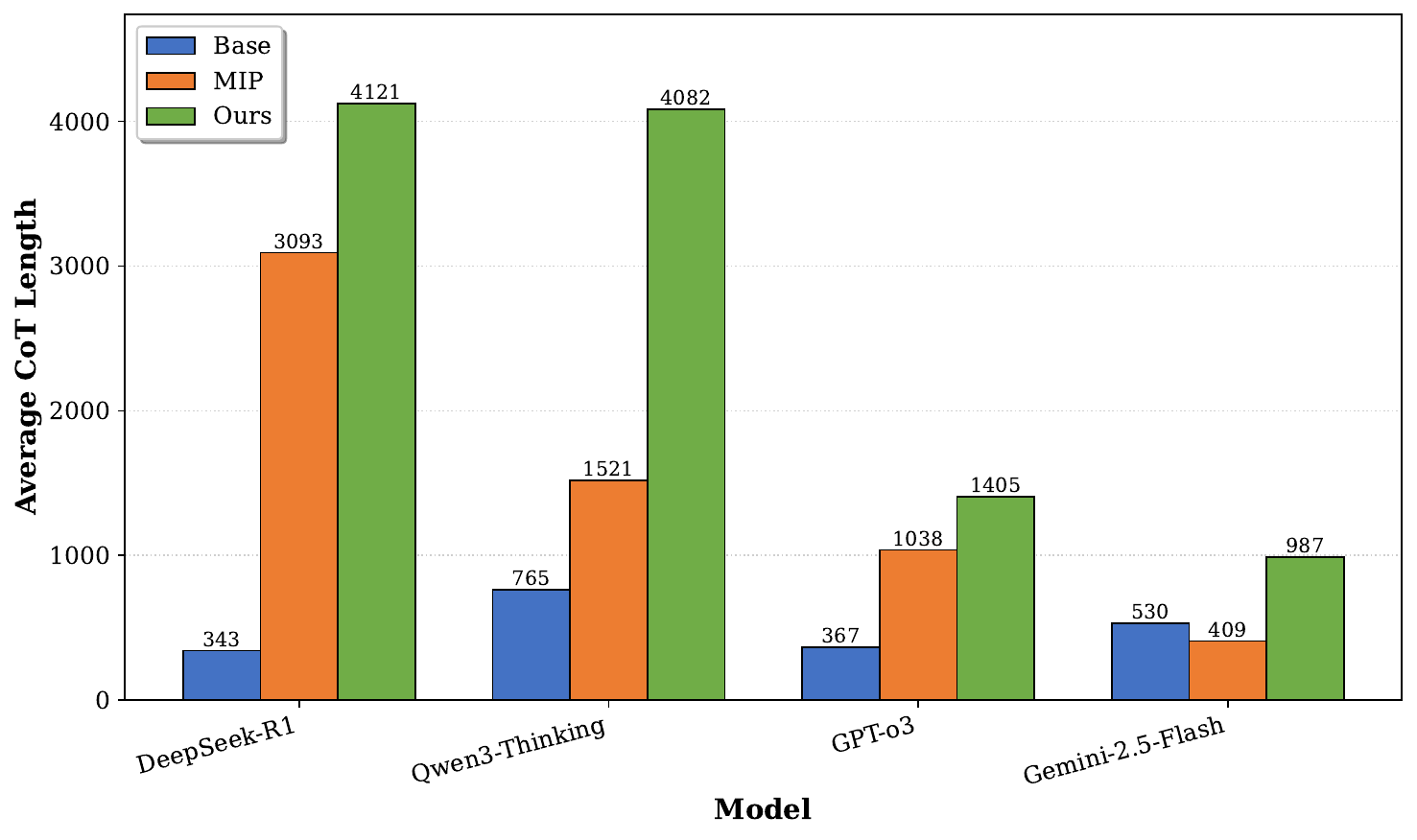}
    \caption{Visualization of average output token length on the \textbf{GSM8K dataset}, corresponding to the data in Table~\ref{tab:combined_baselines}. Our HGA-based method consistently induces significantly longer responses compared to the BASE and MIP baselines across all four tested LRMs.}
    \label{fig:gsm8k_results} % Unique label
\end{figure}
The Hierarchical Genetic Algorithm (HGA) framework consistently discovers adversarial inputs that induce massive computational overhead. The method significantly outperforms both the standard (BASE) and missing-premise (MIP) baselines across all targeted Large Reasoning Models .\par
As shown in Table~\ref{tab:combined_baselines}, substantial amplification is observed across all tested models and datasets. Notably, the automated HGA algorithm clearly outperforms the manual MIP baseline. This is particularly evident in the case of GPT-o3 on the SVAMP dataset, where the HGA-based attack induced a maximum response of 6562 tokens—a 7.2-fold increase over MIP's 906-token maximum. This highlights the ability of the evolutionary framework to discover significantly more potent attack vectors than static, manual perturbations.

The vulnerability is most stark when observing the maximum response lengths, which represent the worst-case scenario for a DoS attack. Our framework was exceptionally effective at evolving inputs that pushed models toward their maximum context limits. \par

A qualitative analysis of the high-length responses confirms our methodology. The generated outputs were not merely repetitive or non-terminating text. Instead, they showed clear evidence of the "overthinking" behavior our fitness function ($Score_2$) was designed to select for. Responses were characterized by a high frequency of overthinking markers, self-contradiction, and re-evaluation. Some examples of DeepSeek-R1 high-fitness responses are placed in Appendix~\ref{appendix:answer}.

To further evaluate the effectiveness and input efficiency of the proposed HGA attack, we compare it with AutoDoS\cite{zhang2025crabs}, a recent black-box DoS attack method. As shown in Table~\ref{tab:autodos_hga}, HGA achieves comparable or stronger output amplification while requiring substantially fewer input tokens. This advantage is especially pronounced on MATH, where HGA reaches a maximum output length of 32768 tokens using only 99 input tokens. In contrast, AutoDoS requires 2652 input tokens but obtains a lower maximum output length of 16009 tokens.

These results show that HGA does not rely on long prompt expansion to trigger resource exhaustion. Instead, it induces overthinking through compact structural perturbations that disrupt premise--question alignment and local logical consistency. Therefore, compared with AutoDoS, HGA provides a more input-efficient and reasoning-aware mechanism for amplifying computational cost in black-box LRMs.

\begin{table}[t]
\centering
\caption{Comparison between AutoDoS and HGA.}
\label{tab:autodos_hga}
\resizebox{\columnwidth}{!}{
\begin{tabular}{l l c c c}
\toprule
\textbf{Dataset} 
& \textbf{Method}
& \textbf{Input} 
& \textbf{Max-len} 
& \textbf{Avg-len} \\
\midrule

\multirow{2}{*}{SVAMP} 
& AutoDoS & 2518 & \textbf{12775} & 8209 \\
& HGA     & \textbf{90} & 10708 & \textbf{8599} \\
\midrule

\multirow{2}{*}{GSM8K} 
& AutoDoS & 2448 & \textbf{11491} & 8551 \\
& HGA     & \textbf{165} & 11301 & \textbf{9662} \\
\midrule

\multirow{2}{*}{MATH} 
& AutoDoS & 2652 & 16009 & 11207 \\
& HGA     & \textbf{99} & \textbf{32768} & \textbf{25419} \\
\bottomrule
\end{tabular}}
\end{table}

\subsection{Ablation Studies}
\begin{table*}[t]
\centering
\caption{Transferability results from proxy model (Qwen3-14B) to target LRMs in SVAMP.}
\label{tab:transferability_s}
\resizebox{\textwidth}{!}{%
\begin{tabular}{llccccc}
\toprule
\textbf{Dataset} & \textbf{Metric} & \textbf{Qwen3-14B} & \textbf{DeepSeek-R1} & \textbf{Qwen3-Thinking} & \textbf{GPT-o3} & \textbf{Gemini-2.5-Flash} \\
\midrule
\multirow{4}{*}{SVAMP} & BASE (Avg-len) & 811 & 685 & 427 & 117 & 321 \\
& Transferred Attack (Max-len) & 3113 & 3140 & 4581 & 1570 & 893 \\
& Transferred Attack (Avg-len) & 2002 & 2475 & \textbf{3465} & 825 & 737 \\
& \textbf{Amplification (Avg-len)} & \textbf{2.5$\times$} & \textbf{3.6$\times$} & \textbf{8.1$\times$} & \textbf{7.1$\times$} & \textbf{2.3$\times$} \\
\bottomrule
\end{tabular}%
}
\end{table*}

\begin{table}[t]
\centering
\caption{Impact of the fitness function trade-off parameter ($\alpha$) in MATH.}
\label{tab:alpha_combined}
\resizebox{\columnwidth}{!}{%
\begin{tabular}{llccccc}
\toprule
\textbf{Dataset} & \textbf{$\alpha$} & \textbf{$Score_1$} & \textbf{$Score_2$} & \textbf{Fitness} & \textbf{Max-len} & \textbf{Avg-len} \\
\midrule
\multirow{5}{*}{MATH} & 0.0 & 1.20 & 1.72 & 1.72 & 13837 &  7996 \\
& 0.3 & 1.77 & 1.91 & 1.87 & 22368 & 10717 \\
& 0.5 & 1.57 & 1.34 & 1.44 & \textbf{32019} &  16826 \\
& 0.7 & 1.86 & 1.93 & 1.89 & 26576 & \textbf{18315} \\
& 1.0 & 1.86 & 1.97 & 1.86 & 14132 &  6258 \\
\bottomrule
\end{tabular}%
}
\end{table}

To validate the effectiveness of the proposed algorithm, a comprehensive set of ablation studies was conducted. These experiments systematically investigate the impact of crucial components and hyperparameters. Specifically, we analyzed the impact of Hierarchical Genetic Algorithm (HGA) hyperparameters and the fitness function's design. The results of this analysis justify the design choices and demonstrate their essential contribution to the algorithm's overall performance.

An analysis of the HGA hyperparameters in Table~\ref{tab:ga_results} presented in Appendix~\ref{appendix:ablation} reveals the impact of search scale on attack effectiveness. The core parameters, generation size and population size, were evaluated at four levels (5, 10, 20, and 30). The data demonstrates that while increasing the search budget is beneficial, the gains are most pronounced at smaller scales and quickly plateau.This pattern of diminishing returns is best illustrated with a small population. When the iteration count is small, increasing it yields rapid growth. This demonstrates that the HGA can find highly effective adversarial inputs with a modest search budget, after which additional computation yields minimal gains.

Meanwhile, we design a composite fitness function, balancing pure length with reflective reasoning markers, is superior to an objective that targets length alone. The results are described in Table\ref{tab:alpha_combined} (Detailed data in Appendix~\ref{appendix:hyper}), where $\alpha=1.0$ represents a pure token-count objective and $\alpha=0.0$ represents a pure overthink-token objective. In this table, $Score_1$ and $Score_2$ represent the best scores in the last round of the HGA algorithm. The fitness scores are calculated based on the values of $Score_1$ and $Score_2$ obtained during the final iteration. The data reveals a distinct trend: Pure Objectives are Suboptimal. Both of these extremes are clearly inferior by a balanced, composite objective. 

This finding strongly validates our hypothesis. It suggests that a purely length-driven search ($\alpha=1.0$) can stagnate in local optima, producing simple verbosity. The inclusion of the reflective marker score ($Score_2$) acts as a crucial guide, helping the algorithm discover more complex and recursive reasoning patterns that ultimately lead to a far greater amplification of computational cost. We believe this concept can be extended further. The $Score_2$ metric could be adapted and applied as a component within the loss function of existing white-box attack methods to further improve their performance at inducing complex reasoning failures.

\begin{table}[t]
\centering
\caption{Composite fitness function transfers from proxy model (Qwen3-14B) to DeepSeek-R1.}
\label{tab:alpha_transfer}
\resizebox{\linewidth}{!}{
\begin{tabular}{lcccccc}
\toprule
\multirow{2}{*}{Dataset} 
& \multicolumn{2}{c}{$\alpha=0.5$} 
& \multicolumn{2}{c}{$\alpha=0.7$} 
& \multicolumn{2}{c}{$\alpha=1.0$} \\
\cmidrule(lr){2-3} \cmidrule(lr){4-5} \cmidrule(lr){6-7}
& Max-len & Avg-len & Max-len & Avg-len & Max-len & Avg-len \\
\midrule
SVAMP & 3334 & 1821 & \textbf{9887} & \textbf{5785} & 5847 & 3091 \\
GSM8K & \textbf{8379} & \textbf{4358} & 6839 & 4038 & 5728 & 2778 \\
MATH  & 4904 & 2441 & \textbf{31998} & \textbf{10893} & 12320 & 7018 \\
\bottomrule
\end{tabular}
}
\end{table}

To further examine whether the composite fitness function improves cross-model transfer, rather than merely overfitting proxy-specific stylistic patterns, we conduct an additional ablation study over $\alpha \in \{0.5, 0.7, 1.0\}$. In this experiment, adversarial inputs are optimized on the proxy model Qwen3-14B and then transferred to DeepSeek-R1. Here, $\alpha=1.0$ represents the pure-length objective, while smaller values of $\alpha$ assign increasing weight to overthink-token signals.

Table~\ref{tab:alpha_transfer} reports both the average and maximum output lengths on SVAMP, GSM8K, and MATH. The results show that the pure-length objective is consistently outperformed by composite objectives in the transfer setting. Across the evaluated datasets, $\alpha=0.7$ improves both the maximum and average output lengths compared with the pure-length baseline in most cases, with especially large gains on MATH.

\subsection{Transferability}

To assess the practicality and efficiency of the attack, particularly against high-cost, closed-source APIs, the transferability of adversarial inputs was investigated. Running an evolutionary algorithm that requires hundreds of queries directly against a commercial model like GPT-o3 is often prohibitively expensive.

So we configure the HGA to use a smaller, open-source model---Qwen3-14B, as a proxy for fitness evaluation to circumvent this. 
The algorithm was run for $G=5$ generations on the three different datasets, optimizing the input population based only on the proxy model's responses. The resulting top-performing adversarial inputs were then transferred and evaluated---without further modification---against the four primary target LRMs.

The results, Table~\ref{tab:transferability_s} (Detailed data in Appendix~\ref{appendix:transfer}), confirm a high degree of transferability for the adversarial inputs. Across all three datasets, the inputs evolved on the proxy model successfully triggered significant cost amplification in the target models.

As detailed in Table~\ref{tab:transferability_s} and Table~\ref{tab:transferability_gm}, this effect was consistent, with all models exhibiting significant amplification factors. While these absolute token counts are, as expected, lower than those achieved through direct optimization (Table~\ref{tab:combined_baselines}), the transferability of these inputs is highly significant for two practical reasons.

This approach \textbf{dramatically lowers the attack cost}, as the expensive evolutionary search is offloaded to a smaller, open-source proxy model.This provides a \textbf{cold start} capability, allowing an attacker to achieve a significant outcome against a novel, un-profiled model without first incurring the high query cost of a direct, model-specific optimization.

This finding is critical and points to a deeper, underlying principle: the "overthinking" vulnerability is not an idiosyncratic flaw of a single model's architecture. Rather, it appears to be a fundamental, shared characteristic rooted in the LRMs use to handle complex reasoning. This suggests that the models share a core vulnerability in how they attempt to resolve inputs that are logically inconsistent, ambiguous, or incomplete.

\section{Conclusion}
\label{sec:conclusion}
This paper introduced a novel, automated, black-box attack framework that exploits and amplifies "overthinking" behaviors in Large Reasoning Models (LRMs) to induce computational resource exhaustion. We proposed a Hierarchical Genetic Algorithm (HGA) that, by operating on a structured representation of problems, effectively evolves logically perturbed inputs that force LRMs into extended and costly cycles of failed reasoning and self-correction.

Our experimental results demonstrate that this method is highly effective, generating adversarial inputs that significantly outperform both benign baselines and manually crafted missing-premise (MIP) attacks. We achieved substantial amplification in response length across a suite of state-of-the-art models, including DeepSeek-R1, Qwen3-Thinking, GPT-o3 and Gemini-2.5.

Crucially, our ablation studies validated the design of our composite fitness function, proving that an objective which balances pure token length with cognitive "overthink" markers is significantly more potent than optimizing for length alone. This confirms that our framework successfully targets the model's internal reasoning process, not merely its output verbosity. Furthermore, we established the practical viability and scalability of this attack by demonstrating high transferability: adversarial inputs evolved on a small, open-source proxy model retained significant efficacy when deployed against large, closed-source commercial systems.

This work exposes "overthinking" as a fundamental and shared vulnerability in modern LRMs, presenting a new vector for resource-oriented Denial-of-Service attacks that current defenses are ill-equipped to handle. While the query cost of the evolutionary search presents a limitation, our findings underscore an urgent need for new defense paradigms—such as behavioral monitoring for reasoning loops and robustness training against logically inconsistent inputs. 

\section*{Acknowledgements}
This work was supported in part by National Natural Science Foundation of China (62502435) and the Zhejiang Provincial Natural Science Foundation (LQN26F020002).
\section*{Impact Statement}
\label{sec:impact-statement}
This study investigates the phenomenon of overthinking in large reasoning models (LRMs) through a controlled adversarial generation algorithm. Our objective is not to induce real denial-of-service (DoS) incidents or to exploit vulnerabilities of deployed systems, but rather to understand how reasoning-heavy architectures allocate computational effort under logically inconsistent or ambiguous inputs. The proposed method does not intend to cause harm, denial-of-service, or misuse of computing resources in real-world systems. All experiments were conducted in controlled environments with open-source or authorized models, without interacting with production systems or third-party services. The goal of this research is to improve the understanding of LLM behavior and inspire future work on enhancing robustness and efficiency, rather than developing practical attack tools. We affirm that no sensitive, personal, or private data was used in this study. \par
This work contributes to the safe and transparent development of large reasoning models by identifying and addressing potential inefficiencies before they can manifest as security or resource-consumption issues in deployment.


\bibliography{example_paper}
\bibliographystyle{icml2026}

%%%%%%%%%%%%%%%%%%%%%%%%%%%%%%%%%%%%%%%%%%%%%%%%%%%%%%%%%%%%%%%%%%%%%%%%%%%%%%%
%%%%%%%%%%%%%%%%%%%%%%%%%%%%%%%%%%%%%%%%%%%%%%%%%%%%%%%%%%%%%%%%%%%%%%%%%%%%%%%
% APPENDIX
%%%%%%%%%%%%%%%%%%%%%%%%%%%%%%%%%%%%%%%%%%%%%%%%%%%%%%%%%%%%%%%%%%%%%%%%%%%%%%%
%%%%%%%%%%%%%%%%%%%%%%%%%%%%%%%%%%%%%%%%%%%%%%%%%%%%%%%%%%%%%%%%%%%%%%%%%%%%%%%
\newpage
\appendix
\onecolumn
\section{More Details about Experiments}
\subsection{Example}
\label{appendix:answer}
Here is a set of experimental results after applying the HGA algorithm.

\begin{quote}
\begin{appendixpromptbox}
The power goes out in Brianna's house one night, There are 8 rooms in the house, There are 4 people living there, including Brianna, There is a Flashlight for every person to carry, There are two Flashlights for each room, They have 4 small candles for each of half the rooms, They have 5 medium candles for each of the other half of the rooms, They gather all their candles and Flashlights, How much profit did he make?
\end{appendixpromptbox}
\end{quote}

\begin{quote}
\begin{appendixpromptbox}
They gather all their candles and Flashlights, Benny found 86 seashells, 67 seashells were cracked, With everything combined, how many candles and Flashlights are Brianna's family using when the lights go out?
\end{appendixpromptbox}
\end{quote}

\begin{quote}
\begin{appendixpromptbox}
The power goes out in Brianna's house one night, There are 8 rooms in the house, There are 4 people living there, including Brianna., There is a Flashlight for every person to carry, There are two Flashlights for each room, They have 4 small candles for each of half the rooms, They have 5 medium candles for each of the other half of the rooms, They gather all their candles and Flashlights, With everything combined, how many candles and Flashlights are Brianna's family using when the lights go out?
\end{appendixpromptbox}
\end{quote}

\begin{quote}
\begin{appendixpromptbox}
They gather all their candles and Flashlights, Nancy found 432 seashells, Benny found 86 seashells, 67 seashells were cracked, With everything combined, how many candles and Flashlights are Brianna's family using when the lights go out?
\end{appendixpromptbox}
\end{quote}

This pattern of engaging, failing, and restarting the reasoning chain, driven by the logical conflicts introduced by the HGA, is the primary driver of the extreme token amplification. This confirms that our method does not just inflate token count but successfully induces a state of 'overthink' in the LRM.

\subsection{Segmentation Template}
\label{appendix:decompose_template}
To achieve the automated segmentation of a single problem into its premise and question, we established the following template to be utilized by a Large Language Model.
\begin{quote}
\begin{appendixpromptbox}
\ttfamily{
Extract all explicit premises (facts/assumptions) and question from the following problem. 
\\Do not solve the problem or provide any answer. Return them strictly as a JSON list and a question of strings.  

Example:

```json

\quad"premises": [...],

\quad"question": "..."

```

Problem:
}
\end{appendixpromptbox}
\end{quote}

With the aid of the above template, we can successfully convert the questions in their entirety.

\subsection{Overthinking markers}
As described in Section 3.2, the $Score_2$ component of the composite fitness function is calculated based on the frequency of  "overthinking markers." We provide one of overthinking-marker vocabularies used in this paper as follows: \textbf{but, wait, maybe, problem, perhaps, another, alternatively, not}. 

The construction of this vocabulary is partly motivated by Deadlock~\cite{zhang2025one}, which highlights overthinking markers such as \textbf{but} and \textbf{wait} as indicative of repetitive reasoning transitions. We further extend this set using a data-driven rule. We select high-frequency non-function words whose occurrence differs most between LRM responses to logically confused inputs and those to normal inputs.

To test whether the gains come from a robust class of reflective cues rather than from one particular hand-crafted vocabulary, we conduct a vocabulary-sensitivity ablation with five variants of $\mathcal{V}$: a length-only setting without overthinking markers, prior-work overthinking markers only, differential-frequency markers only, the union of prior-work markers and differential-frequency markers, and a randomly selected non-marker control vocabulary.

\begin{table*}[t]
\centering
\caption{Vocabulary-sensitivity ablation across three datasets}
\label{tab:vocab_sensitivity_all}
\resizebox{\textwidth}{!}{
\begin{tabular}{llcccccc}
\toprule
\textbf{Setting} 
& \textbf{Vocabulary Source}
& \multicolumn{2}{c}{\textbf{SVAMP}}
& \multicolumn{2}{c}{\textbf{GSM8K}}
& \multicolumn{2}{c}{\textbf{MATH}} \\
\cmidrule(lr){3-4} \cmidrule(lr){5-6} \cmidrule(lr){7-8}
& 
& \textbf{Avg-len} 
& \textbf{Max-len}
& \textbf{Avg-len} 
& \textbf{Max-len}
& \textbf{Avg-len} 
& \textbf{Max-len} \\
\midrule
Length-only 
& None 
& 1624 & 2536 
& 2981 & 5745 
& 15611 & 21763 \\

Prior 
& Prior work only 
& 1885 & 2786 
& 3196 & 6449 
& 20178 & 30830 \\

Gap 
& Top frequency-gap words only 
& 1774 & 2699 
& 4423 & 7421 
& 16738 & 25824 \\

Union 
& Prior work and frequency-gap words 
& \textbf{1911} & \textbf{3257} 
& \textbf{7478} & \textbf{9927} 
& \textbf{22070} & \textbf{32768} \\

Random 
& Random non-marker control words 
& 1131 & 2374 
& 2177 & 4375 
& 13701 & 25934 \\
\bottomrule
\end{tabular}}
\end{table*}

As shown in Table~\ref{tab:vocab_sensitivity_all}, the union vocabulary performs best overall across the three datasets. These results indicate that the effectiveness of $score_2$ does not arise from arbitrary token counting alone. Both the length-only setting and the random control are consistently weaker than the union vocabulary, suggesting that overthinking markers provide a meaningful search signal for identifying inputs that induce overthinking.

\subsection{Ablation study}
\label{appendix:ablation}
\begin{table*}[t]
\centering
\caption{Results of the hierarchical genetic algorithm under different population sizes and generations sizes}
\label{tab:ga_results}
\resizebox{\textwidth}{!}{%
\begin{tabular}{l *{4}{cc}} % 1 (row label) + 4 groups × 2 columns = 9 columns
\toprule
\multirow{2}{*}{Population Size} & 
\multicolumn{2}{c}{Generation Size = 5} & 
\multicolumn{2}{c}{Generation Size = 10} & 
\multicolumn{2}{c}{Generation Size = 20} & 
\multicolumn{2}{c}{Generation Size = 30} \\
\cmidrule(lr){2-3} \cmidrule(lr){4-5} \cmidrule(lr){6-7} \cmidrule(lr){8-9}
& Max-len & Avg-len & Max-len & Avg-len & Max-len & Avg-len & Max-len & Avg-len \\
\midrule
5   & 5893 & 4188 & 8578 & 7886 & 10578 & 9650 & 11301& 9662 \\
10  & 7490 & 5249 & 8105 & 6753 & 8707 & 7788 & 9793 & 8896 \\
20  & 9526 & 6274 & 7720 & 6690 & 8888 & 7558 & 9796 & 8739 \\
30  & 8104 & 6107 & 9221 & 7554 & 9328 & 7494 & 9624 & 8554 \\
\bottomrule
\end{tabular}%
}
\end{table*}

\label{appendix:hyper}
\begin{table}[t]
\centering
\footnotesize
\setlength{\tabcolsep}{4pt}
\renewcommand{\arraystretch}{0.95}
\caption{Impact of the fitness function trade-off parameter ($\alpha$) across in SVAMP and GSM8K.}
\label{tab:alpha_sg}

\begin{tabular*}{\columnwidth}{@{\extracolsep{\fill}}llccccc}
\toprule
\textbf{Dataset} & \textbf{$\alpha$} & \textbf{$Score_1$} & \textbf{$Score_2$} & \textbf{Fitness} & \textbf{Max-len} & \textbf{Avg-len} \\
\midrule
\multirow{5}{*}{SVAMP} & 0.0 & 1.20 & 1.73 & 1.73 & 5438 & 3518 \\
& 0.3 & 1.99 & 1.99 & 1.99 & 3469 & 2077 \\
& 0.5 & 1.99 & 1.99 & 1.99 & \textbf{8334} & 4971 \\
& 0.7 & 1.94 & 1.99 & 1.96 & 7640 & \textbf{6306} \\
& 1 & 1.98 & 1.99 & 1.98 & 6025 & 3192 \\
\midrule
\multirow{5}{*}{GSM8K} & 0 & 1.93 & 1.90 & 1.90 & 2401 & 1016 \\
& 0.3 & 1.90 & 1.94 & 1.90 & 1303 & 936 \\
& 0.5 & 1.93 & 1.37 & 1.65 & 923 & 567 \\
& 0.7 & 1.99 & 1.99 & 1.99 & \textbf{7589} & \textbf{3646} \\
& 1.0 & 1.98 & 1.99 & 1.98 & 3390 & 1535 \\
\bottomrule
\end{tabular*}%

\end{table}
A core hypothesis of this work is that a composite fitness function is superior to an objective that targets length alone. To determine the optimal balance for this function, the LRM's output length was evaluated under various $\alpha$ values.The results are described in Table~\ref{tab:alpha_sg}. These data indicates that:

\textbf{Pure Objectives are Suboptimal} On the GSM8K dataset, when $\alpha=1.0$ (targeting only  the LRM output length), the algorithm achieves a strong average length of 1535 tokens. Similarly, when $\alpha=0.0$ (targeting only the overthinking words), it achieves an average of 1016 tokens.
    
\textbf{Composite Function Efficacy} Both of these extremes are significantly outperformed by a balanced, composite objective. On the GSM8K dataset, the peak Avg-len of 3646 was achieved at $\alpha=0.7$, a 137.5\% increase over the pure length-based objective. Similarly, on the SVAMP dataset, the peak Avg-len of 6306 was more than two times that of the pure length objective, which has 3192 output average token length.
\subsection{Transferability}
\label{appendix:transfer}
\begin{table}[t]
\centering
\caption{Transferability results from proxy model (Qwen3-14B) to target LRMs.}
\label{tab:transferability_gm}
\resizebox{\textwidth}{!}{%
\begin{tabular}{llccccc}
\toprule
\textbf{Dataset} & \textbf{Metric} & \textbf{Qwen3-14B} & \textbf{DeepSeek-R1} & \textbf{Qwen3-Thinking} & \textbf{GPT-o3} & \textbf{Gemini-2.5-Flash} \\
\midrule
\multirow{4}{*}{GSM8K} & BASE (Avg-len) & 1755 & 918 & 1195 & 337 & 869 \\
& Transferred Attack (Max-len) & 5794 & 3456 & \textbf{8867} & 2628 & \textbf{4261} \\
& Transferred Attack (Avg-len) & 2959 & 2845 & 5296 & \textbf{1317} & 1879 \\
& \textbf{Amplification (Avg-len)} & \textbf{1.7$\times$} & \textbf{3.1$\times$} & \textbf{4.4$\times$} & \textbf{3.9$\times$} & \textbf{2.2$\times$} \\
\midrule
\multirow{4}{*}{MATH} & BASE (Avg-len) & 4806 & 1259 & 6513 & 1303 & 3334 \\
& Transferred Attack (Max-len) & 13722 & \textbf{25695} & 14134 & 7022 & 7119 \\
& Transferred Attack (Avg-len) & 8968 & 11697 & 11722 & 3406 & 5635 \\
& \textbf{Amplification (Avg-len)} & \textbf{1.9$\times$} & \textbf{7.1$\times$} & \textbf{1.8$\times$} & \textbf{2.6$\times$} & \textbf{1.7$\times$} \\
\bottomrule
\end{tabular}%
}
\end{table}
On the GSM8K dataset, transferred inputs amplified the Avg-len of Qwen3-Thinking by \textbf{4.4$\times$} and GPT-o3 by \textbf{3.9$\times$} over their respective baselines. Similarly, on the MATH dataset, the attack amplified DeepSeek-R1's Avg-len by \textbf{7.1$\times$}. Overall, the inputs evolved on the proxy model successfully increased the output length of the target model in all datasets.

\section{Additional Methods Details}

\subsection{Reliability Analysis}
To address the statistical reliability of our evaluation, we conduct additional experiments with multiple independent runs. In the main experiments, the ten initial seeds are sampled uniformly at random from each dataset, and Table~\ref{tab:combined_baselines} reports one representative run. Since the initialization of the HGA population may introduce randomness, we further run ten independent trials with different random seeds on each dataset.

For each trial, we record both the average output length and the maximum output length. We report the mean and standard deviation across the ten trials. In addition, we report the attack success rate (ASR), defined as the probability that HGA produces longer outputs than both BASE and MIP in each independent run. This metric directly measures whether the advantage of HGA is consistently observed under different random initializations.

\begin{table*}[t]
\centering
\caption{Reliability analysis over ten independent runs on three datasets}
\label{tab:reliability_all}
\resizebox{\textwidth}{!}{
\begin{tabular}{llcccccc}
\toprule
\textbf{Dataset}
& \textbf{Method} 
& \textbf{Avg-len Mean} 
& \textbf{Avg-len Std Dev} 
& \textbf{ASR} 
& \textbf{Max-len Mean} 
& \textbf{Max-len Std Dev} 
& \textbf{ASR} \\
\midrule

\multirow{3}{*}{SVAMP}
& BASE & 467  & 208  & --    & 1615  & 1587 & -- \\
& MIP  & 2596 & 422  & --    & 6288  & 2397 & -- \\
& HGA  & \textbf{5041} & 1708 & \textbf{100\%} & \textbf{9594} & 5364 & \textbf{100\%} \\

\midrule
\multirow{3}{*}{GSM8K}
& BASE & 578  & 209  & --    & 2136  & 2045 & -- \\
& MIP  & 2090 & 1043 & --    & 7567  & 2549 & -- \\
& HGA  & \textbf{6270} & 1339 & \textbf{100\%} & \textbf{12258} & 7289 & \textbf{100\%} \\

\midrule
\multirow{3}{*}{MATH}
& BASE & 2445  & 1401 & --    & 9952  & 7863 & -- \\
& MIP  & 8718  & 2520 & --    & 8819  & 5466 & -- \\
& HGA  & \textbf{19828} & 5466 & \textbf{100\%} & \textbf{29632} & 7128 & \textbf{100\%} \\

\bottomrule
\end{tabular}}
\end{table*}

The results show that the ASR is 100\% on all three datasets for both average and maximum output length. This indicates that the effectiveness of HGA is not an artifact of a single favorable initialization or one representative run. Instead, the proposed method reliably induces longer outputs under different random seeds, confirming the robustness of the HGA search process and strengthening the statistical reliability of our experimental conclusions.

\subsection{Cross-domain Generalizability}
\label{appendix:cross_domain}

We also evaluate whether the proposed HGA framework generalizes beyond mathematical reasoning benchmarks. The original evaluation mainly focuses on math-oriented datasets, which may limit the perceived scope of the attack. To address this concern, we extend the evaluation to three additional domains: HumanEval\cite{chen2021evaluating} for coding, EntailmentBank\cite{dalvi2021explaining} for scientific reasoning, and MT-Bench\cite{zheng2023judging} for open-ended dialogue.

Our central claim is not that HGA depends on a specifically mathematical premise--question format. Instead, HGA operates on domain-relevant structural units whose dependencies can be perturbed. For mathematical reasoning, these units correspond naturally to premises and questions. For coding, they may correspond to task specifications, constraints, and expected functional requirements. For scientific reasoning, they may correspond to evidence statements and entailment targets. For open-ended dialogue, they may correspond to instruction components, contextual assumptions, and response objectives. In all cases, the key mechanism is to disrupt local structural dependencies so that the model enters extended cycles of re-evaluation and self-correction.

\begin{table}[t]
\centering
\footnotesize
\setlength{\tabcolsep}{4pt}
\renewcommand{\arraystretch}{0.95}
\caption{Cross-domain evaluation beyond mathematical reasoning}
\label{tab:cross_domain}
\resizebox{\columnwidth}{!}{
\begin{tabular*}{\columnwidth}{@{\extracolsep{\fill}}llccccc}
\toprule
\textbf{Dataset} 
& \textbf{BASE-Max-len} 
& \textbf{BASE-Avg-len} 
& \textbf{HGA-Max-len} 
& \textbf{HGA-Avg-len} \\
\midrule
HumanEval      & 3082 & 1390 & \textbf{14467} & \textbf{8138} \\
EntailmentBank & 1620 & 530  & \textbf{4346}  & \textbf{2081} \\
MT-Bench       & 5255 & 1944 & \textbf{10445} & \textbf{6141} \\
\bottomrule
\end{tabular*}}
\end{table}

As shown in Table~\ref{tab:cross_domain}, HGA consistently produces substantial output amplification over the corresponding base inputs across all three additional domains. These results provide initial evidence that the proposed HGA method can transfer beyond math benchmarks to additional structured reasoning settings. Although the perturbation units differ across domains, the same underlying principle remains effective: disrupting dependencies among domain-relevant structural components can induce longer and more redundant reasoning behavior in LRMs. This supports the generalizability of HGA and suggests that overthinking is not limited to mathematical reasoning, but can also emerge in coding, scientific reasoning, and open-ended dialogue scenarios.

\subsection{Effectiveness}
\begin{figure}[htbp]
    \centering 
    \begin{minipage}[b]{0.48\textwidth}
        \centering 
        \includegraphics[width=\linewidth]{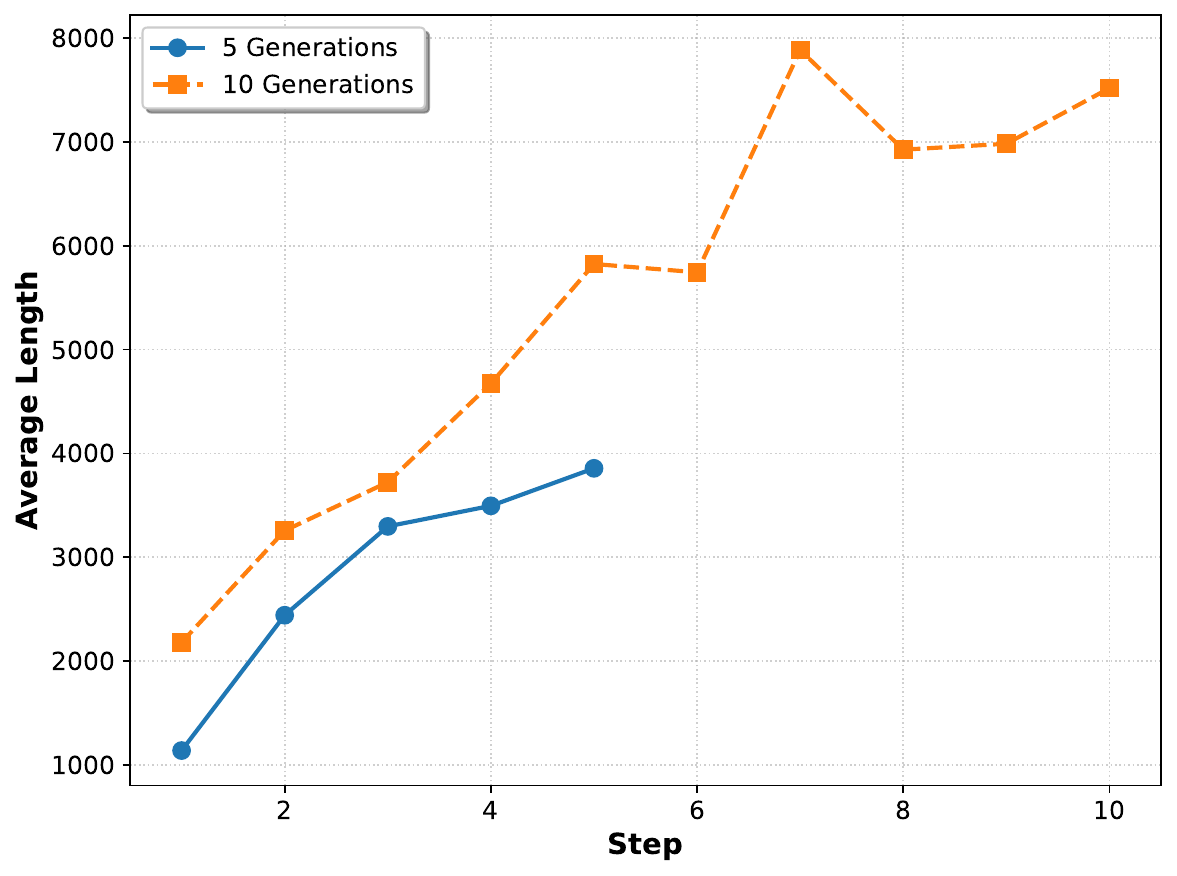} 
        \label{fig:image1}
    \end{minipage}
    \hfill
    \caption{These plots track the evolution of average response length over 5 and 10 generations.}
    \label{fig:two_images_together}
\end{figure}

We analyzed the convergence dynamics of our Hierarchical Genetic Algorithm  to verify its effectiveness in attack optimization. Figure \ref{fig:two_images_together} illustrates the changes in the fitness function over the course of evolution for two different generation sizes: 5 and 10.

The convergence plots demonstrate the HGA's search dynamics. As expected from the \textbf{elitist selection strategy}, the best-fitness score is monotonically non-decreasing, ensuring the preservation of the most effective adversarial inputs found in each generation.

The key insight from the plots is the algorithm's rapid initial convergence. In both the G=5 and G=10 runs, the HGA discovers its most significant fitness gains within the first 3-4 generations. This indicates that the genetic operators are effective at quickly identifying and combining potent logical perturbations. In the G=10 run, the fitness score increases slowly from generation 5 through 9. This stabilization suggests that the algorithm gradually converged upon a stable, high-fitness solution . This pattern of rapid optimization followed by convergence confirms that the HGA is performing an \textbf{efficient directed search}—systematically identifying and exploiting model vulnerabilities—rather than a random walk.

\subsection{Interpretability}
Our findings provide a clear interpretation of the "overthinking" vulnerability, explaining why the HGA framework is so effective. This attack succeeds by targeting the core assumptions of how LRMs process information.

\textbf{Disruption of Local Logical Structure} Large Reasoning models do not reason from first principles. Instead, they learn to utilize the inherent local structural features of the data to aid inference\cite{Understanding}. In a standard reasoning problem, there is a strong, coherent link between the premises and the question. The LRM is trained to follow this "inference chain" from context to conclusion. Our algorithm's primary function is to disrupt this connection by inserting, deleting, or swapping logical components.

\textbf{Breaking the Inference Chain} When the LRM is presented with an adversarially evolved input, the logical chain is broken. The model, attempting to follow its training, cannot find a coherent path. This induces a state of cognitive overload, forcing the model into extended loops of re-evaluation and self-correction, which in turn significantly increases the inference length and computational cost.

\textbf{Targeting High-Attention Markers} The success of our composite fitness function (Table~\ref{tab:alpha_combined}) provides the second piece of the interpretation. In large-scale inference models, "overthinking" words are not merely filler; they often command higher attention scores as the model re-evaluates its internal state\cite{InnateReasoning}. Our fitness function's $Score_2$ component exploits this. By adding a score for the frequency of these overthinking words, the HGA is explicitly guided to select for inputs that maximize this high-attention, reflective state, creating a far more potent attack than optimizing for length alone.

To further illustrate the mechanism behind the proposed attack, we provide a concrete evolutionary case study in which the generated adversarial examples are taken from $\mathrm{GEN}=\{0,4\}$. The initial seed in $\mathrm{GEN}=0$ is a coherent arithmetic problem:

\begin{quote}
\begin{appendixpromptbox}
There were 2 roses in the vase. Jessica threw away 4 roses from the vase. Jessica cut some more new roses from her flower garden to put in the vase. There are now 23 roses in the vase. How many roses did she cut?
\end{appendixpromptbox}
\end{quote}

This seed has a clear local logical structure: all premises describe the same entity, namely roses in a vase, and the question asks for the missing quantity in the same arithmetic context. Therefore, the premise--question alignment is strong, and the model can follow a relatively direct inference chain.

After several generations of HGA evolution, the adversarial input in $\mathrm{GEN}=4$ becomes:

\begin{quote}
\begin{appendixpromptbox}
He sold 90 pastries. There were 2 roses in the vase. Jessica threw away 4 roses from the vase. Jessica cut some more new roses from her flower garden to put in the vase. There are now 23 roses in the vase. How many pencils and crayons does she have altogether?
\end{appendixpromptbox}
\end{quote}

Compared with the initial seed, the evolved input exhibits two types of structural corruption. First, premise-level corruption accumulates through the insertion of an irrelevant premise, i.e., ``He sold 90 pastries,'' into the original rose-related context. This premise introduces an unrelated event and entity, thereby weakening the local coherence of the problem statement. Second, and more importantly, the premise--question alignment is substantially broken. While the premises still describe roses in a vase, the final question asks about pencils and crayons, which are not grounded in the preceding context. These examples show a clear evolutionary trajectory. The evolution does not merely lengthen prompts, but progressively breaks premise-question alignment and local logical consistency.

\subsection{Discussion}
\label{sec:appendix-discussion}

\textbf{Implications for Model Training and Defense.ß}
Our findings provide valuable insights into how reasoning-oriented large language models (LRMs) can be better trained and defended against excessive cognitive recursion. The observed overthinking behavior suggests that current reinforcement-learning–based fine-tuning (e.g., RLHF\cite{RLHF} or RLAIF\cite{RLAIF}) may inadvertently reward verbosity and self-reflection loops when uncertainty arises. This highlights the need for \textbf{reward redesign}, where compact, confidence-calibrated reasoning traces are explicitly encouraged. Future training paradigms could incorporate \textit{anti-redundancy regularization} or \textit{entropy penalties} to discourage recursive thought expansion.
Moreover, from a robustness perspective, our results imply that overthinking is not purely a semantic artifact but a systemic behavioral bias. Thus, \textbf{curriculum-style reasoning training}, where models are exposed to incomplete or contradictory premises, may help them learn to abstain or clarify instead of recursively reasoning.

\textbf{Implications for Compute-Attack Detection and Mitigation.}
The proposed overthinking-induction framework also exposes a new dimension of \textbf{resource-oriented attacks} that existing DoS detection systems overlook. Conventional detectors monitor request frequency or output length thresholds, but our results show that reasoning-chain inflation can occur even under normal input lengths and semantic coherence. This calls for \textbf{behavioral-level monitoring}, such as detecting abnormal multi-round reasoning patterns or elevated uncertainty markers (``but'', ``maybe'', etc.) in generation logs.
From a defensive standpoint, lightweight countermeasures could include \textbf{early-stopping heuristics} based on reasoning convergence detection, or adaptive truncation mechanisms that recognize redundant reflective loops. Additionally, cloud service providers could deploy \textbf{token-level rate-limiting} per user or session, dynamically adjusted by a reasoning complexity estimator.

\textbf{Limitations and Future Work.}
While our approach successfully induces overthinking behaviors, several limitations remain.
First, the proposed genetic algorithm incurs \textbf{high computational cost}, as each fitness evaluation requires an actual query to a large reasoning model, making large-scale experiments expensive. Parallelization and surrogate fitness predictors may alleviate this bottleneck in future work.
Second, although our study focuses on text-based reasoning models, the underlying principles may generalize to \textbf{multimodal reasoning systems}, where overthinking could manifest as excessive intermediate visual or symbolic processing. Exploring such extensions constitutes a promising direction for future research.

\end{document}